\def\tomath#1{\ifmmode{#1}\else $#1$\fi}%
\def\eV{\ifmmode{eV}\else$eV$\fi}
\def\keV{\ifmmode{keV}\elsekeV\fi}
\def\MeV{\ifmmode{MeV}\elseMeV\fi}
\def\GeV{\ifmmode{GeV}\else$GeV$\fi}
\def\sr{\ifmmode{\,sr}\else$sr$\fi}
\def\hfq{\hfill\quad}
\def\cc#1{\hfq#1\hfq}
\def\ccc#1{\hfill$\,$#1$\,$\hfill}
\def\hfl#1#2{\smash{\mathop{\hbox to 12mm{\rightarrowfill}}\limits^{\scriptstyle#1}_{\scriptstyle#2}}}
\def\*{\hbox{\hphantom{(0)}}}
\def\0{\hbox{\hphantom{0}}}
\def\1{\hbox{\hphantom{\MeV}}}
\def\noy#1#2{\tomath{{}^{#1}#2}}
\def\tvi{\vrule height 12pt depth 5pt width 0pt}
\def\tv{\tvi\vrule}

\def\placefigure#1#2#3{%
\begin{figure}%
\centerline{\psfig{figure=#2.eps,width=#1cm}}\smallskip%
\caption[]{#3}%
\protect\label{#2}%
\end{figure}}%
\def\rotatefigure#1#2#3#4{%
\begin{figure}[h]%
\centerline{\psfig{figure=#3.eps,width=#1cm,angle=#2}}\smallskip%
\caption[]{#4}%
\protect\label{#3}%
\end{figure}}%
\def\noy#1#2{\tomath{{}^{#1}}#2}
\documentclass[epj]{svjour}
\usepackage{psfig}
\begin{document}

\title{Decay of proton-rich nuclei between \noy{39}{Ti} and \noy{49}{Ni}}
\author{J.~Giovinazzo\inst{1} \and B.~Blank\inst{1} \and C.~Borcea\inst{2} \and
M.~Chartier\inst{1} \and S.~Czajkowski\inst{1} \and
G.~de~France\inst{3} \and 
R.~Grzywacz\inst{4}\thanks{\emph{present address: Department of Physics and Astronomy, 
University of Tennessee, Knoxville, TN 37996-1200, USA}} \and Z.~Janas\inst{4} \and  M.~Lewitowicz\inst{3} \and 
F.~de~Oliveira Santos\inst{3} \and M.~Pf\"utzner\inst{4} \and M.S.~Pravikoff\inst{1} \and  J.C.~Thomas\inst{1}}

\institute{Centre d'Etudes Nucl\'eaires de Bordeaux-Gradignan, Le Haut-Vigneau, 
B.P. 120, F-33175 Gradignan Cedex, France
\and
Institute of Atomic Physics, P.O. Box MG6, Bucharest-Margurele, Romania
\and 
Grand Acc\'el\'erateur National d'Ions Lourds, B.P. 5027, F-14076 Caen Cedex, France
\and
Institute of Experimental Physics, University of Warsaw, PL-00-681 Warsaw, Hoza 69, Poland
}

\date{Received: date / Revised version: date}

%
%
%
\abstract{
Decay studies of very neutron-deficient nuclei
ranging from \noy{39}{Ti} to \noy{49}{Ni} have been performed during a
projectile fragmentation experiment at the GANIL/LISE3 separator.
For all nuclei studied in this work, \noy{39,40}{Ti}, \noy{42,43}{Cr},
\noy{46}{Mn}, \noy{45,46,47}{Fe} and \noy{49}{Ni}, half-lives and decay
spectra have been measured. In a few cases, $\gamma$ 
coincidence measurements helped to successfully identify the initial 
and final states of transitions. In these cases, partial decay scheme 
are proposed. For the most exotic isotopes, \noy{39}{Ti}, \noy{42}{Cr}, 
\noy{45}{Fe} and \noy{49}{Ni}, which are candidates for
two-proton radioactivity from the ground state, no clear evidence of this
process is seen in our spectra and we conclude rather on a delayed particle 
decay.
\PACS{27.40.+z \and 21.10.Dr \and 23.50.+z \and 25.70.Mn
} 
} 

\maketitle
%
%
%
\section{Introduction}

Up to the $A = 50$ mass region, the proton drip-line has been reached
experimentally.
The observation of nuclei very close to this limit of existence was
possible due to recent developments of projectile fragmentation
facilities.
Indeed, in the mass region we are interested in here, a recent experiment
at the GSI/FRS facility allowed for the observation of \noy{42}{Cr}~($T_z=-$3) 
as well as of \noy{45}{Fe} and \noy{49}{Ni}~($T_z=-7/2$)~\cite{blank95kr}.
During an experiment at the GANIL/LISE3 separator, the existence of these isotopes
has been confirmed with increased statistics and \noy{48}{Ni} has been
observed for the first time~\cite{blank00ni48}.

When going close to the proton drip-line, the emission of two protons
from a nuclear state becomes possible.
This decay mode is mainly sequential if the emission of one proton can occur to
an intermediate state.
This process has first been observed in the $\beta 2p$ decay of
\noy{22}{Al}~\cite{cable83}.
Other two-proton emissions from an excited state have been studied, mainly
in the region from $A=20$ to 40~\cite{cable84,jahn85,axelsson98,borrel87,aysto85},
but no evidence of a non-sequential decay could be pointed out.

If there is no accessible intermediate state, the sequential emission is forbidden,
and the emission of the two protons is simultaneous.
This situation mainly expected to occur from nuclear ground states.
Then two cases are possible.
Firstly, the decay may proceed via a three-body desintegration as observed
for \noy{12}{O}~\cite{kryger95} or \noy{6}{Be}~\cite{bochkarev92}.
Secondly, the nuclear state may decay via \noy{2}{He} emission and a
strong angular and energy correlation
between the two resulting protons may be expected.
Up to now, this second process has never been observed experimentally.

In the $A\simeq 50$ mass region, a specific interest arises for decay 
studies since several nuclei in this region are candidates for this correlated
emission of two protons from the ground state:
\noy{39}{Ti}, \noy{42}{Cr}, \noy{45}{Fe},
\noy{48,49}{Ni}~\cite{brown91,ormand97,cole96}.
In addition, full shell-model calculations in the $fp$-shell are now
possible~\cite{caurier94} and may allow for a valuable comparison
with experimental results to improve our understanding of nuclear structure
in this region and to deduce effective interactions.

In the present experiment, we have studied the decay of the
isotopes \noy{39,40}{Ti}, \noy{42,43}{Cr},
\noy{46}{Mn}, \noy{45,46,47}{Fe} and \noy{49}{Ni}, for which only few
experimental information was available. The nucleus studied in most detail 
prior to our experiment is \noy{40}{Ti}~\cite{bhattacharya98,liu98} which
will be used as a reference for calibration.
For \noy{39}{Ti}, the half-life and the main decay branches, including the
$\beta 2p$ decay, have been observed previously~\cite{detraz90,moltz92}.
Some information, like rough lifetime estimates and a few delayed
proton lines, is also available for
\noy{43}{Cr}, \noy{46}{Mn} and \noy{46,47}{Fe}~\cite{borrel92}.
For the most exotic isotopes, \noy{42}{Cr}, \noy{45}{Fe} and \noy{49}{Ni},
only their existence was reported~\cite{blank95kr}.

In the present work, we take advantage of the delayed proton emission to
determine a precise half-life of the isotopes of interest.
Many transitions by proton emission are observed, but only few of
them could be placed in a partial decay scheme.
For the most abundantly produced isotopes, some $\gamma$ lines have been registered
in coincidence with protons yielding valuable information on the
decay schemes.
Clearly identified transitions also allowed us to determine the position
of the isobaric analog state (IAS) in some daughter nuclei and thus, by 
means of the isobaric multiplet mass equation (IMME), to get an estimate 
for the mass excess of the exotic emitter nuclei.

%
%
%
\section{Experimental procedure}

Neutron deficient nuclei have been produced in the fragmentation reaction
of a ${}^{58}$Ni$^{26+}$ primary beam at 74.5 $MeV/nucleon$ from the GANIL
facility.
The 230.6~$mg/cm^2$ thick natural nickel production target and the 2.7~$mg/cm^2$
carbon stripper were located between the solenoids of the SISSI
device\cite{sissi}.
The most exotic isotopes have been selected with the Alpha spectrometer
and the LISE3~\cite{lise} separator, including a shaped degrader 
(10.4~$mg/cm^2$ of beryllium) at the intermediate focal plane and a
Wien filter at the end of the LISE3 beam line.

The selected isotopes have been implanted in a silicon telescope.
The telescope consisted (see figure~\ref{e312a_setup})
in two silicon detectors ($\Delta{}E1$ and $\Delta{}E2$) for energy-loss
measurements (300~$\mu m$ each, 600 and $30\times 30$ $mm^2$),
the second one being position sensitive.
The ions were implanted in a third silicon diode, $E3$ (600~$mm^2$ surface and
300~$\mu m$ thickness).
Two additional silicon detectors ($E4$ with 450~$mm^2$ surface and 700~$\mu m$
thickness and $E5$ with 600~$mm^2$ surface and 6~$mm$ thickness)
were used to veto
particles which were not stopped in the previous diodes.
The use of two veto detectors is usually not required. However, it was
initially planned to implant the ions of interest in the fourth detector 
and to use only the last one as a veto.
This procedure had to be changed in order to correctly identify $^{48}$Ni with
a maximum of parameters (see~\cite{blank00ni48}).
Therefore, no trigger was properly set up for radioactivity events
in the E3 detector.
Radioactivity events were triggered only by $\beta$ particles
in the second and fourth detector leading to an efficiency of
$(38\pm 3) \%$ (see below).
For each of the detectors E2, E3, and E4, two electronic chains with different
amplification gains were used for implantation and radioactivity events.

The isotope identification was achieved by energy-loss and residual-energy
measurements in the telescope and by time of flight measurements. Details of the
identification procedure are given in ref.~\cite{blank00ni48}.

\placefigure{8}{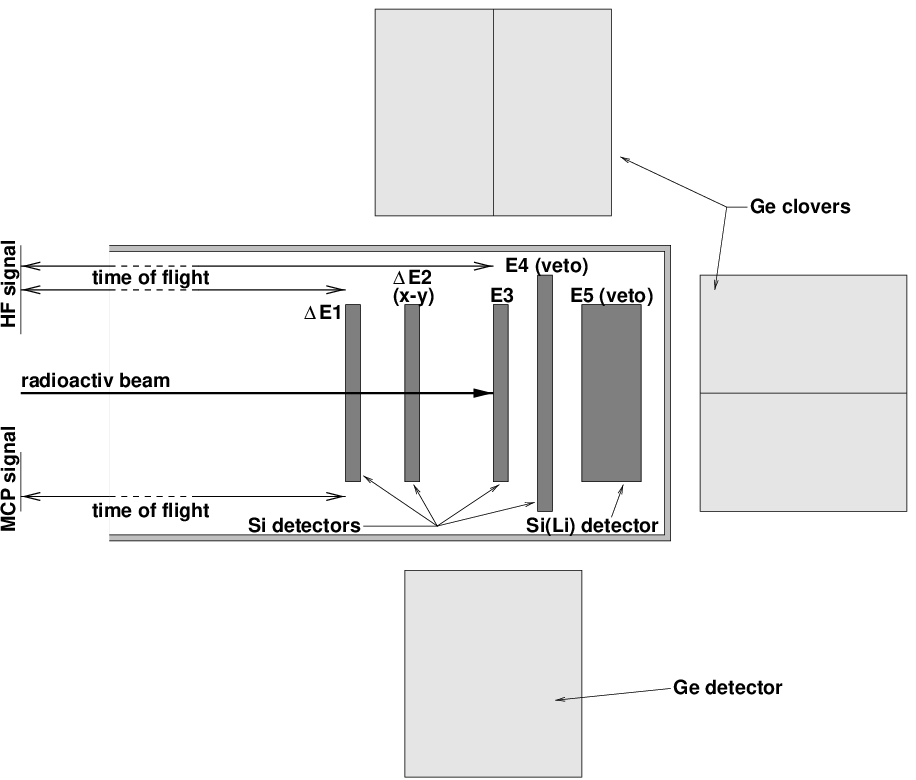}{Detection setup: Ions are implanted in the $E3$ 
           detector. The identification of isotopes is performed with the
           $\Delta E1$ and $\Delta E2$ energy loss information, several 
           time-of-flight measurements, the total energy and position 
           measurements. For details see ref.~\cite{blank00ni48}.}

This setup was surrounded by three germanium detectors, a single cristal
and two four-cristal clovers. The total efficiency was measured to be 
1.12\% at 1.17 MeV. 
The $\gamma$ detectors were only used for coincidences and did not appear
in the trigger logic of the acquisition.

%
%
\section{Data analysis}

As described in the paper on the $^{48}$Ni observation~\cite{blank00ni48}, 
the identification of isotopes is performed on the basis of a
total-energy versus energy-loss matrix. For this purpose,
the low-gain electronic chains for the $E1$, $E2$, and $E3$ detectors 
were calibrated by means of simulations with the LISE~\cite{lise_code_new} 
program. Additionnal conditions were imposed on
the position in the $E2$ detector, the different times of flight, and
the signals in the veto detector.

A clock with a 100 $\mu s$ step was used for time correlations between
radioactivity and implantation events. In the analysis, 
each radioactivity event following an implantation is taken into account 
as long as no further implantation did occur.
Although the implantation rate was about 10 ions per second in the 
$E3$ detector, most of those isotopes lie much closer to stability 
(with long half-lifes) and do not emit delayed protons.
Consequently, the correlation procedure is stopped only after the
implantation of a relatively exotic nucleus defined by a large window
around nuclei of interest in the $\Delta E - E$ matrix. 
Simulations have been performed using the experimental time distributions
$i)$ for all nuclei and $ii)$ for the most exotic ones only.
They clearly demonstrated that the determination of half-lives lower
than 100~ms do not require any correction for the cut introduced
by following implantations.
A correction would be needed, on the contrary, if this cut were performed
by any implantation instead of the most exotic ones only.

In order to reduce the background from the decay of long-lived isotopes
accumulated in the implantation detector, we will analyse proton energy
distributions with a condition on events above 1 MeV.
This cut removes almost all $\beta$ events that represent the main
background, while most of the proton strength is distributed at
higher energies.

The fits for half-life determination are performed with the maximum-likelyhood
method from the PAW/MINUIT software~\cite{paw}.
When the proton emission feeds a nucleus that does not decay by proton
emission, $i.e.$ there is no contamination from protons of the
daughter decay in the energy spectrum, an exponential function with
a constant background is used for the fit.
This concerns the decays of \noy{40,39}{Ti}, \noy{43}{Cr}, \noy{46}{Mn} and
\noy{47}{Fe}.
For the other isotopes, \noy{42}{Cr}, \noy{46,45}{Fe} and \noy{49}{Ni},
the delayed proton emission from the daughter nucleus is included
in the fitting procedure. The error bars include statistical and
systematic errors. The systematic errors are mainly due to the time window 
used for the fitting procedure and to the proton energy cut.
These systematic errors were determined by changing both the time and
energy cuts.
They vary between 20\% and 50 \% of the total error.

The proton energy calibration is obtained from the peaks at 1702, 2162 and
3733 keV in the decay of \noy{40}{Ti}~\cite{liu98,bhattacharya98}.
Due to the summing of proton and $\beta$ particles energies, the proton 
peaks have to be fitted with a Gaussian and an exponential tail plus an
exponential background due to lower-energy proton groups.
\noy{40}{Ti} together with \noy{43}{Cr} is also used to estimate the
trigger efficiency for radioactivity events after implantation.
This efficiency is estimated from the fit of the decay time distribution:
The number of counts for events with an energy above 1~MeV is compared to 
the number of implantations. With the additional information that,
for \noy{40}{Ti}, the total proton branching ratio above 1 MeV is
$(97\pm 2) \%$~\cite{liu98}, we deduce a detection efficiency 
$\epsilon_P=(39\pm 2)\%$ from the decay of this nucleus.
Due to the high $Q_{EC}$
value ($\sim 16\ $MeV) and the low proton separation energy of only about 
200 keV in \noy{43}{V}, we assume that $(95\pm 5) \%$ of the decay 
of \noy{43}{Cr} proceeds by delayed particle emission. 
The resulting detection efficiency for radioactive
events is then $\epsilon_P=(37\pm 2) \%$.
In the following, we will use a mean value of $\epsilon_P = (38\pm 3) \%$
for the trigger efficiency for radioactivity events.

%
%
\section{Results}

\subsection{The decay of ${}^{40}$Ti}

As mentionned above, \noy{40}{Ti} is used as a reference since its decay
has already been studied extensively~\cite{bhattacharya98,liu98}.
The proton-energy and decay-time distributions are presented in figure
\ref{ti40_all}.
For energy calibration, peaks 2, 3 and 4 have been fitted.
The relative intensities are reported in table~\ref{ti40} and
are in good agreement with previous experiments.
In the present experiment, the $\gamma$ detection efficiency was too low 
to observe the 2469~keV line in \noy{39}{Ca} in coincidence with 
the 1330 keV~proton peak mentioned by Liu {\it et al.}~\cite{liu98}
and Bhattacharya {\it et al.}~\cite{bhattacharya98}.

\placefigure{8}{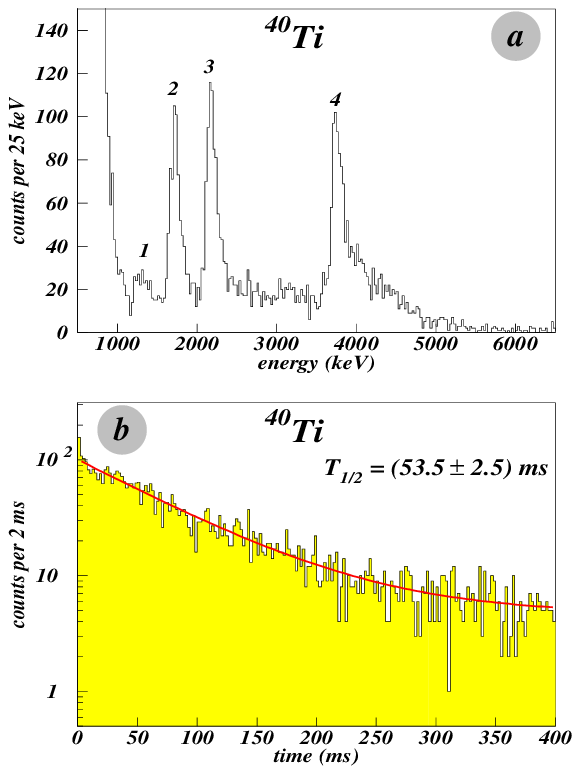}{$a)$ Decay-energy distribution of
           \noy{40}{Ti}. The main proton lines are reported in table \ref{ti40} 
           for comparison with previous work.
           $b)$ Decay-time distribution of decay events after
           implantation of \noy{40}{Ti}. The deduced half-life is 
           ($53.5\pm 2.5)\ $ms (see table~\ref{periode}).}

The half-life is estimated by fitting the distribution with an
exponential decay and a constant background, since there is no
contribution from the decay of the daughter nucleus.
The resulting value of $T_{1/2} = (53.5 \pm 2.5)\ $ms is in good agreement
with previous measurements and close to
theoretical predictions (see table \ref{periode}).

\begin{table*}
\centerline{\vbox{\offinterlineskip
\halign{\tvi\tv\ccc{#}&\tv\ccc{#}&\ccc{#}\tv&\ccc{#}&\ccc{#}\tv&\ccc{#}&\ccc{#}\tv\cr
\noalign{\hrule}
peak &\multispan2{\tvi\tv\ccc{this work}}\tv&\multispan2{\ccc{Liu {\it et al.} 
\cite{liu98}}}\tv&\multispan2{\ccc{Bhattacharya {\it et al.} \cite{bhattacharya98}}}\tv\cr
number &$E_p$ (keV)  &$b_P$ (\%) &  $E_p$ (keV)  &$b_P$ (\%) &$E_p$ (keV)  &$b_P$ (\%) \cr
\noalign{\hrule}
1   &1332 (25)  &\05.8 (20) &1322 (25) &\04.35 (82) &1325 (7)\0 &\03.58 (61) \cr
2   &1703 (10)  &21.7 (30)  &1705 (10) &22.5\0 (21) &1701 (6)\0 &23.8\0 (6) \cr
3   &2154 (9)\0 &27.4 (36)  &2167 (10) &28.5\0 (19) &2160 (6)  &29.8\0 (7) \cr
4   &3736 (14)  &23.8 (36)  &3731 (10) &22.8\0 (19) &3734 (7)  &21.7\0 (5) \cr
\noalign{\hrule}
}}}
\caption{Branching ratios of main proton lines in the decay of \noy{40}{Ti}. 
         The results of the present work are compared to data from 
         literature.}
\label{ti40}
\end{table*}

\subsection{The decay of ${}^{39}$Ti}

The limited statistics of the present work as well as of the work
from D\'etraz {\it et al.}~\cite{detraz90} and Moltz {\it et al.}~\cite{moltz92}
does not allow for a detailed comparison of the energy spectra.
However, from the proton spectrum in figure \ref{ti39_all}a, it seems to be
likely that the peak at $(4880\pm 40)\ $keV (marked as 4)
corresponds to the $\beta 2p$ decay to the ground state of \noy{37}{K} as
already observed by D\'etraz {\it et al.} and Moltz {\it et al}.
This energy is 130~keV higher than in the work of
Moltz {\it et al.}~\cite{moltz92} and leads to an excitation energy of the 
isobaric analogue state (IAS) in \noy{39}{Sc} of $(8.96 \pm 0.06)\ $MeV.
From this result, we deduce a mass excess for this analog state 
of $\Delta m = (-5210 \pm 85)\ $keV. Using the mass excess of the 
\noy{39}{Cl} ground state ($\Delta m = -29799\ $keV) and of 
the \noy{39}{Ar} IAS ($\Delta m = -31334\ $keV),
the mass excess resulting from the isobaric multiplet mass equation ($IMME$)
for \noy{39}{Ti} is
$\Delta m = 1790\pm 90\ $keV, with $IMME$ coefficients in agreement
with what is expected from a calculation with a uniformly charged
sphere~\cite{benenson79}.

\placefigure{8}{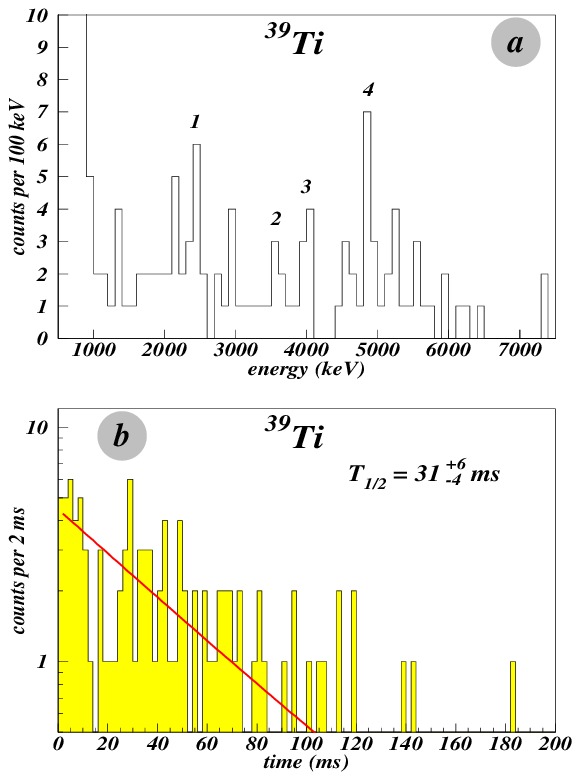}{
          $a)$ Decay-energy distribution from the decay of
          \noy{39}{Ti}. The peak labels correspond to those used in 
          table~\ref{ti39}.
          $b)$ Decay-time distribution of decay events after
          implantation of \noy{39}{Ti}. The fit yields a half-life of 
          $T_{1/2} = 31^{+6}_{-4}\ $ms.}

From this result, it follows that \noy{39}{Ti} is unbound by 670~keV with
respect to two-proton emission. This is in nice agreement 
with several theoretical $Q_{2p}$ values given
in table~\ref{ti39_q2p}. When we use the mass excess of the IAS in \noy{39}{Sc}
as determined in the present work to calculate the coefficients of the $IMME$
for the $A=39, T_z=-5/2$ nuclei, we get coefficients which are somewhat closer
to the result expected from a uniformly charged sphere (the expected ratio
is 1, our value is $1.1\pm 0.2$) as when using 
the result of Moltz {\it et al.} ($1.2\pm 0.2$). 
However, as the error bars are large, no final conclusion is possible.

Although other proton groups are visible in the spectrum of
figure~\ref{ti39_all}a, 
no further transition can be clearly identified
to support or reject the assignment of the two-proton transition. The most prominent 
charged-particle groups are summarized in table~\ref{ti39}
with their relative intensity.

\begin{table}
\centerline{\vbox{\offinterlineskip
\halign{\tvi\tv\cc{#}&\cc{#}\tv\cr
\noalign{\hrule}
reference    &$S_{2p}$ (keV)\cr
\noalign{\hrule}
this work with IMME                  & $-670\pm 100$ \cr
Moltz {\it et al.}~\cite{moltz92}      & $-530\pm 65$ \cr
Brown~\cite{brown91}                   & $-657$ \cr
Cole~\cite{cole96}                     & $-798\pm 39$ \cr
Ormand~\protect\cite{ormand96}                & $-666\pm 107$ \cr
\noalign{\hrule}
}}}
\caption{Comparison of two-proton separation energies for \noy{39}{Ti}
         from the present work with theoretical predictions and with the 
         experimental datum from Moltz {\it et al}.}

\label{ti39_q2p}
\end{table}

As \noy{39}{Sc} is unbound, the decay of \noy{39}{Ti} proceeds with a 100\%
branching ratio via delayed-particle emission.
From figure \ref{ti39_all}b, we deduce a half-life of
$T_{1/2} = 31^{+6}_{-4}\ $ms for \noy{39}{Ti} which is in good agreement
with predicted values and previous experimental work~\cite{detraz90}
(see table~\ref{periode}).
There is no contamination in the proton-energy distribution from the decay
of \noy{38}{Ca} ($\beta p$ daughter) or \noy{37}{K} ($\beta 2p$ daughter) 
which are much longer lived and do not decay via proton emission.

\begin{table}
\centerline{\vbox{\offinterlineskip
\halign{\tvi\tv\cc{#}&\cc{#}&\cc{#}\tv\cr
\noalign{\hrule}
peak num    &energy (keV) &branching (\%)\cr
\noalign{\hrule}
1   &2440 (25) &\08.0 (50) \cr
2   &3575 (30) &\06.5 (45) \cr
3   &3990 (30) &\07.3 (45) \cr
4   &4880 (40) &12.5 (65) \cr
\noalign{\hrule}
}}}
\caption{Branching ratios and decay energies of the main proton lines in 
        the decay of
        \noy{39}{Ti}. Peak 4 is suggested to correspond to the $\beta$2p decay
        to the ground state in \noy{37}{K} via the IAS in \noy{39}{Sc}.}
\label{ti39}
\end{table}

\subsection{The decay of ${}^{43}$Cr}

The proton-energy distribution of the decay of \noy{43}{Cr} is shown in
figure~\ref{cr43_all}a. 
Using the mass evaluation of Audi and Wapstra~\cite{audi97}, we assign the 
charged-particle peak labelled 9 to the $\beta 2p$ decay via the IAS in \noy{43}{V}
to the ground state in \noy{41}{Sc}. This interpretation is supported by
the fact that this peak has a somewhat larger width than those for other transitions
which might be explained by the different relative angles between the two 
protons yielding different recoil energies for the heavy fragment.
As the energy loss of the heavy recoil is subject to a strong pulse 
height defect~\cite{ratkowski75}, two-proton lines are usually broadened. From this 
information, the excitation energy of the IAS in \noy{43}{V} 
is deduced to be $(8250\pm 25\pm 230)\ $keV, where the second uncertainty
comes from the \noy{43}{V} mass extrapolation.

\placefigure{8}{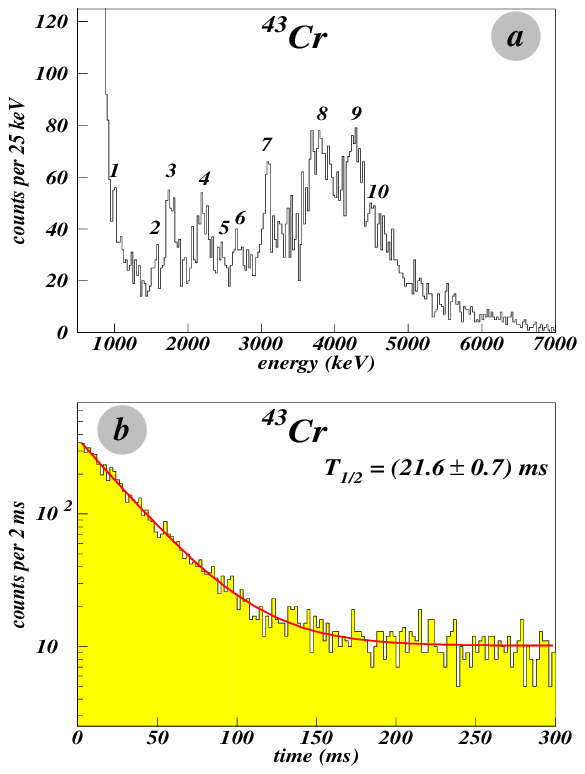}{$a)$ Decay-energy distribution in the decay of
          \noy{43}{Cr}. The peak labels correspond to those in table \ref{cr43}.
           Peak 9 is tentatively identified as originating from a $\beta$2p 
           transition via the IAS in \noy{43}{V}. 
           $b)$ Decay-time distribution of decay events after
           implantation of \noy{43}{Cr}. A half-life of (21.6$\pm$0.7) ms is
           deduced.}

The $\gamma$-energy distribution (see fig. \ref{cr43_gamma}) in coincidence
with protons shows a strong $\gamma$ line at $(1553.5\pm 1.0)\ $keV.
As shown in figure \ref{cr43_gamma}a, this $\gamma$ peak shows up in 
coincidence with all charged-particle peaks all over the decay-energy 
distribution. As this $\gamma$ ray de-excites the first excited state in
\noy{42}{Ti}, we conclude on only a very small proton branching to
the ground state of \noy{42}{Ti}.

\placefigure{8}{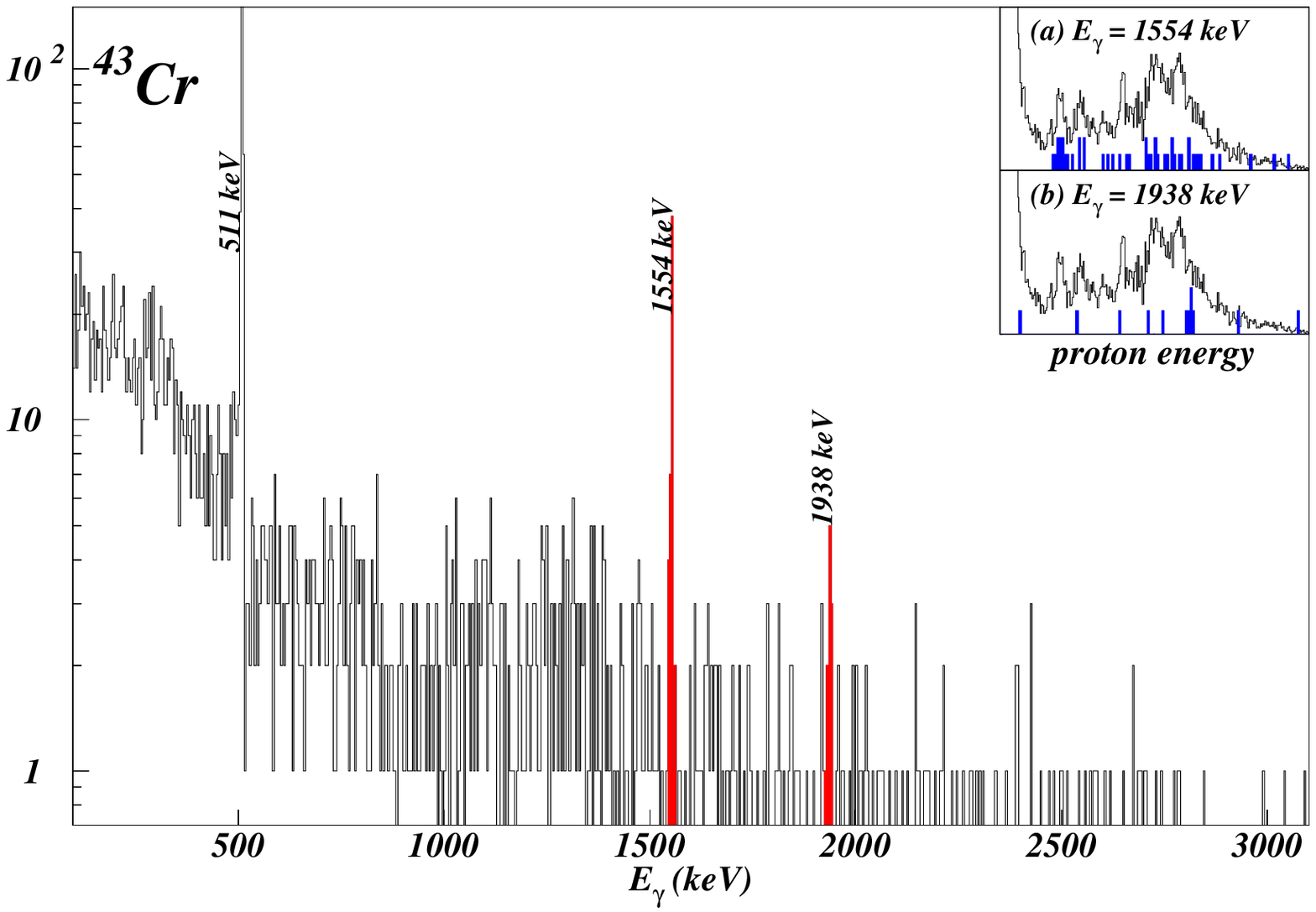}{$\gamma$-energy spectrum as measured in 
        coincidence with charged particles. The insert a) shows the  
        charged-particle events in coincidence with the 1554~keV $\gamma$ 
        line. It evidences that 
        this $\gamma$ line is in coincidence with almost all 
        charged-particle groups. The 1938~keV line is mainly in 
        coincidence with proton peak 10 (insert $b$).}

Another line at $(1938\pm 2)\ $keV is seen in coincidence with the 
charged-particle peak 10 at $E = 4590\ $keV (see fig. \ref{cr43_gamma}b).
Such a $\gamma$ transition is neither known in \noy{42}{Ti} ($\beta$p daughter)
nor in \noy{41}{Sc} ($\beta$2p daughter). If we assume that the $\beta$ feeding
above the IAS in \noy{43}{V} is weak, the $4590\ $keV peak can not originate
from a $\beta$2p decay, as the Q value for a transition from the IAS to
the \noy{41}{Sc} ground state is only about 4~MeV. A consistent picture
arises if we assume that the $4590\ $keV line links  the IAS
to an excited state in \noy{42}{Ti} at 3492~keV. This state would decay 
by a $\gamma$ cascade of the 1938~keV and the 1554~keV $\gamma$-rays.
Such a level is not known in
\noy{42}{Ti}, but this might be explained by the fact that this nucleus
is only poorly studied. In the mirror nucleus $^{42}$Ca, a $I^\pi = 3^-$ level
is known at 3446.96~MeV which indeed decays by a cascade of $\gamma$ rays of 
1922.18~keV and 1524.73~keV. However, if the assumed spins (see 
figure~\ref{cr43_decay}) are correct, such a transition would be a $\ell = 1$
transition and should be suppressed.
Nevertheless, such a hypothesis leads to an excitation energy 
of the IAS in \noy{43}{V} of $(8270\pm 45\pm 230)\ $keV
($\pm 230\ $keV from the \noy{43}{V} mass uncertainty) in good agreement
with the one deduced from the $\beta 2p$ transition as mentioned above.
The energy resulting from both estimates is then
$E^{*}_{IAS} = (8255\pm 25\pm 230)\ $keV.
With this information, we propose the partial decay scheme shown in 
figure~\ref{cr43_decay}.

\begin{table}
\centerline{\vbox{\offinterlineskip
\halign{\tvi\tv\cc{#}&\cc{#}&\cc{#}\tv\cr
\noalign{\hrule}
peak num    &energy (keV) &branching ratio (\%)\cr
\noalign{\hrule}
1   &1008 (17) &\02.0 (6)\0 \cr
2   &1563 (18) &\02.3 (7)\0 \cr
3   &1780 (16) &\06.0 (12) \cr
4   &2222 (16) &\06.4 (13) \cr
5   &2500 (21) &\03.3 (9)\0 \cr
6   &2717 (22) &\04.0 (10) \cr
7   &3078 (16) &\07.7 (12) \cr
8   &3828 (23) &18.3 (21) \cr
9   &4292 (22) &14.8 (2.8) \cr
10  &4590 (45) &\03.8 (20) \cr
\noalign{\hrule}
}}}
\caption{Branching ratios and decay energies of the main
        charged-particle groups in the decay of \noy{43}{Cr}. 
        The labels correspond to those used in figure~\ref{cr43_all}a.}
\label{cr43}
\end{table}

The half-life of \noy{43}{Cr} was determined in the usual way with a condition
on the charged-particle spectrum to select events with a decay energy above
1~MeV. We deduce a half-life of $T_{1/2} = (21.6\pm 0.7)\ $ms
(see fig. \ref{cr43_all}b). This value is in agreement with but much more precise
than the value from Borrel {\it et al.}~\cite{borrel92} of 27$^{+32}_{-10}$ms.

\placefigure{8}{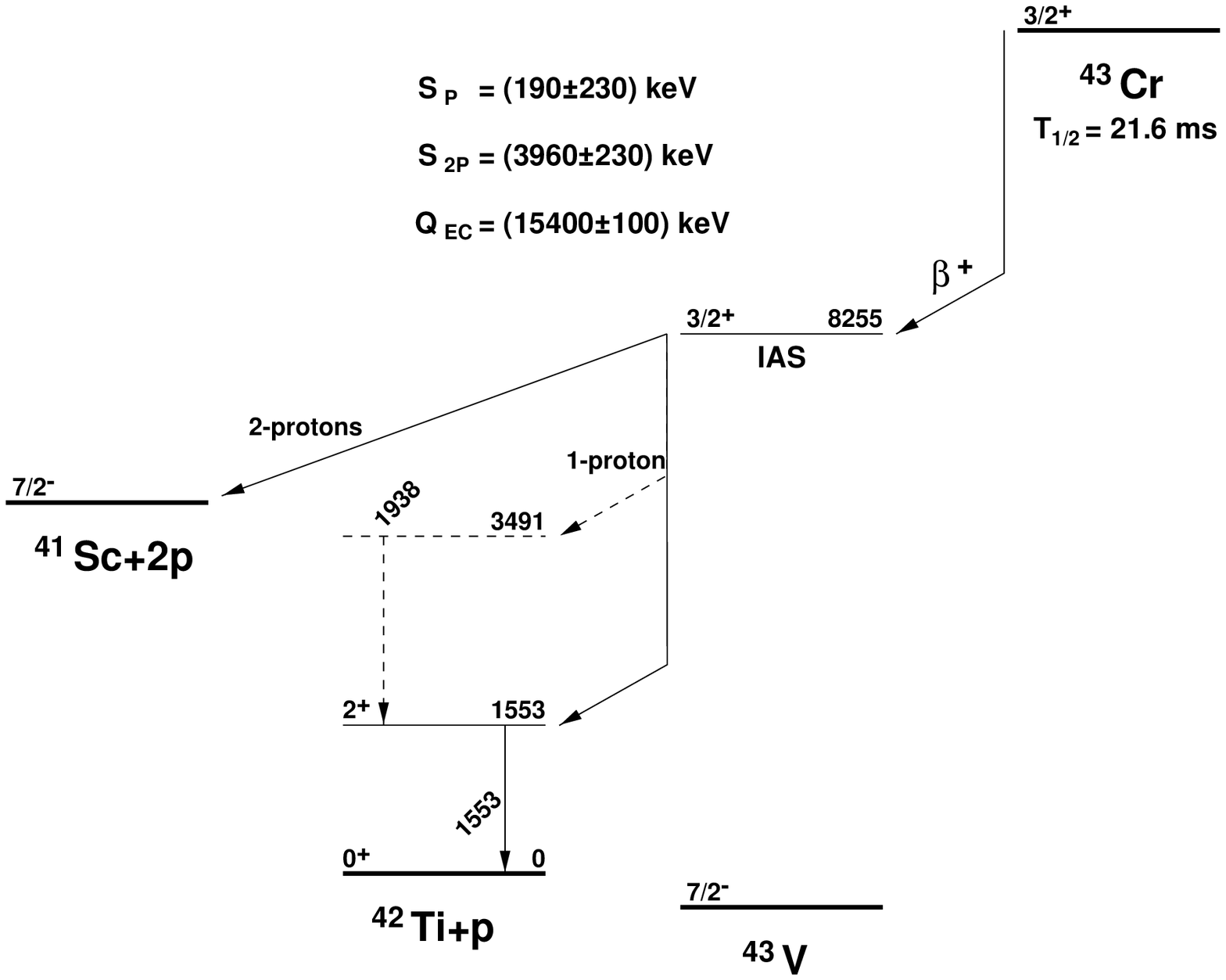}{Partial decay scheme for \noy{43}{Cr} as deduced
             from the present work.}

\subsection{The decay of ${}^{42}$Cr}

The isotope \noy{42}{Cr} is a possible candidate for two-proton 
radioactivity. However, according to all theoretical 
predictions~\cite{brown91,ormand97,cole96}, its $Q_{2p}$ value is probably too 
small for a detectable two-proton branching ratio. 
Our experimental results are in agreement with this expectation.
When inspecting the charged-particle spectrum in figure~\ref{cr42_all}a,b, 
the most prominent peak is the line at 1.9~MeV. If this peak were due to
two-proton radioactivity, the barrier-penetration half-life of \noy{42}{Cr}
would be of the order of $10^{-12}$s, orders of magnitude lower than the 
flight time from the production target to the detector, and this isotope 
would decay on its flight through the separator (see figure~\ref{cr42_mass}). 
The observation of \noy{42}{Cr} 
in our experiment and in previous work rejects several models predicting
\noy{42}{Cr} highly unbound with $Q_{2p}$ values of 1~MeV and higher.

\placefigure{8}{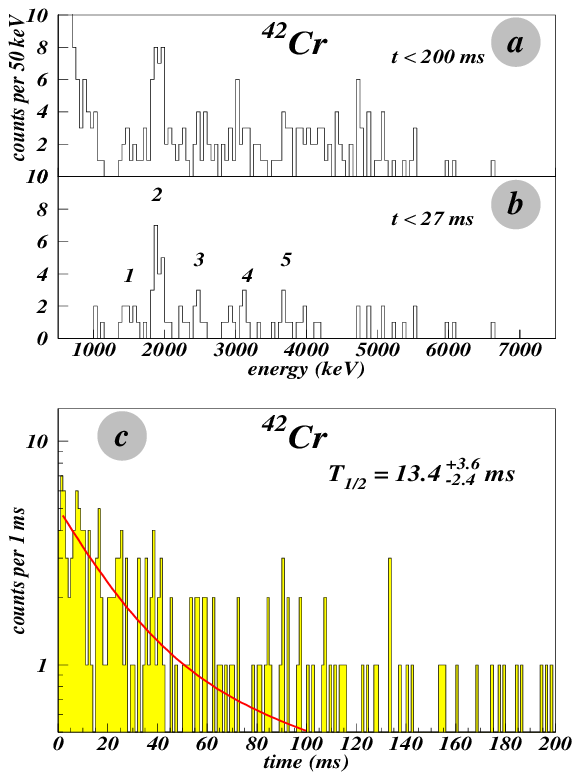}{$a,b)$ Decay-energy distribution in the decay of
         \noy{42}{Cr}. The histogram in part $a)$ shows the decay energy for 
         events decaying up to 200~ms after a \noy{42}{Cr} implantation, 
         whereas the histogram in $b)$ is due to events up to 27~ms
         corresponding to two half-lives. 
         The short time cut favors mainly events from the short-lived 
         \noy{42}{Cr} (see table~\ref{cr42}), while the daughter decay 
         contribute to the 200~ms histogram.
         $c)$ Time distribution of decay events after
         implantation of \noy{42}{Cr}. The fit includes the contribution
         of the decay of \noy{41}{Ti}.}

Another hint against a two-proton decay of \noy{42}{Cr} is the half-life 
determined for charged-particle events with an energy above 1~MeV 
(figure~\ref{cr42_all}c). Our experimental data yield a half-life of
$T_{1/2} = 13.4^{+3.6}_{-2.4}\ $ms. Such a half-life is typical for a 
$\beta$ decay in this region. We therefore conclude that \noy{42}{Cr}
is decaying mainly by $\beta$-delayed charged-particle emission and that
a possible two-proton radioactivity is very weak.

\rotatefigure{8}{-90}{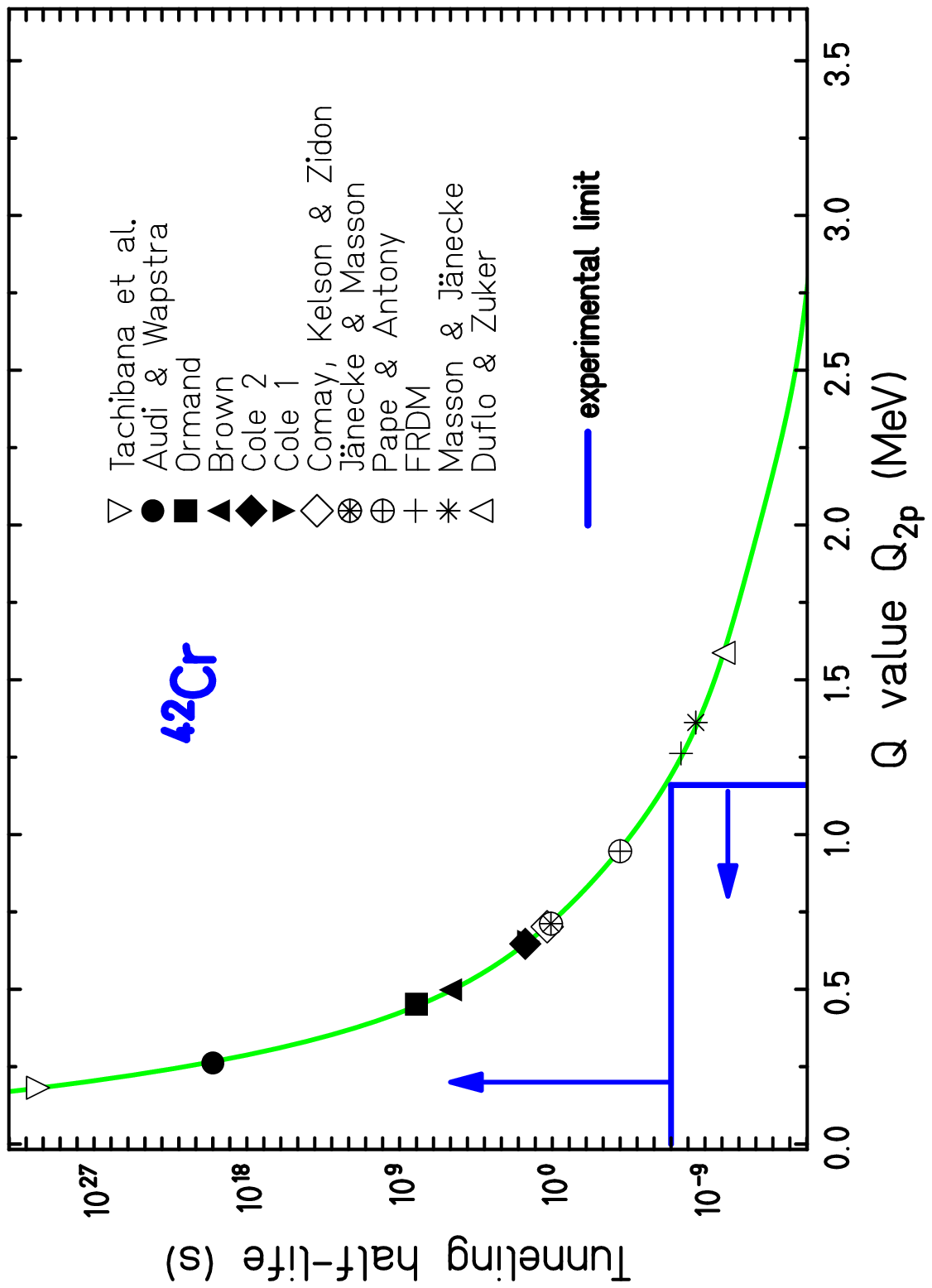}{%
       Barrier-penetration half-life of \noy{42}{Cr}
       for the tunneling of a di-proton through the Coulomb barrier
       as a function of the two-proton separation energy. $Q_{2p}$ values
       resulting from different mass
models~\cite{masses88,audi97,brown91,ormand97,cole96,moeller95,duflo95,tachibana90} are shown.}

Each $\beta$ decay of \noy{42}{Cr} is followed by a proton emission
since the daughter nucleus, \noy{42}{V}, is unbound.
In the case of a $\beta p$ transition, the decay is followed
by another $\beta p$ emission from \noy{41}{Ti} with a half-life of
$(82\pm 3)\ $ms~\cite{liu98}.
In the proton spectrum of figure \ref{cr42_all}a,b, two cuts at $t=200\ $ms 
and $t=27\ $ms (corresponding to two half-lives) have been performed 
in order to discriminate contributions from \noy{41}{Ti} and \noy{42}{Cr}.
From this analysis, we conclude that the charged-particle group at 1.9~MeV
is due to the decay of \noy{42}{Cr}. However, due to too low statistics and
therefore no coincident $\gamma$ rays, we can not attribute this peak to
a well defined transition.

Besides the peak at 1.9~MeV, no other pronounced structure appears in the
spectrum. The transition strength located around 4 to 5 MeV is probably due
to the daughter decay of \noy{41}{Ti}~\cite{liu98}. According to the structure 
of the mirror nucleus of \noy{40}{Sc}, the 2490 keV charged-particle 
line could be the $\ell = 0$ two-proton transition from
the IAS to the first excited $0^+$ state in \noy{40}{Sc} which, according to 
its mirror \noy{40}{K} lies at around 1.6~MeV, but the statistics is too 
low to observe a coincident $\gamma$ ray.

\begin{table}
\centerline{\vbox{\offinterlineskip
\halign{\tvi\tv\cc{#}&\cc{#}&\cc{#}\tv\cr
\noalign{\hrule}
peak number &energy (keV) &branching ratio (\%)\cr
\noalign{\hrule}
1   &1500 (35) &\09.2 (73)\0 \cr
2   &1905 (20) &29.2 (100) \cr
3   &2490 (30) &\09.2 (73)\0 \cr
4   &3110 (20) &\07.7 (42)\0 \cr
5   &3715 (20) &\06.2 (40)\0 \cr
\noalign{\hrule}
}}}
\caption{Branching ratios and decay energies of the main charged-particle 
         groups in the decay of
         \noy{42}{Cr}. The peak numbers correspond to those used in 
         figure~\ref{cr42_all}b.}
\label{cr42}
\end{table}

\subsection{The decay of ${}^{46}$Mn}

Fig.~\ref{mn46_all}a shows the decay-energy distribution of \noy{46}{Mn}. 
The expected positions for $\beta$p and $\beta$2p transitions via the IAS 
in \noy{46}{Cr} to the ground states of \noy{45}{V} and \noy{44}{Ti}
are indicated as determined from the mass evaluation of Audi and 
Wapstra~\cite{audi97}, but no peak can be clearly identified.
The spectrum rather resembles the typical bell-shaped proton distribution
which is obtained when a high number of states is populated which
then decay by proton emission (see e.g.~\cite{giovinazzo00}). Much 
higher statistics is needed to locate decay-energy peaks linked for example
to the decay of the IAS.

\placefigure{8}{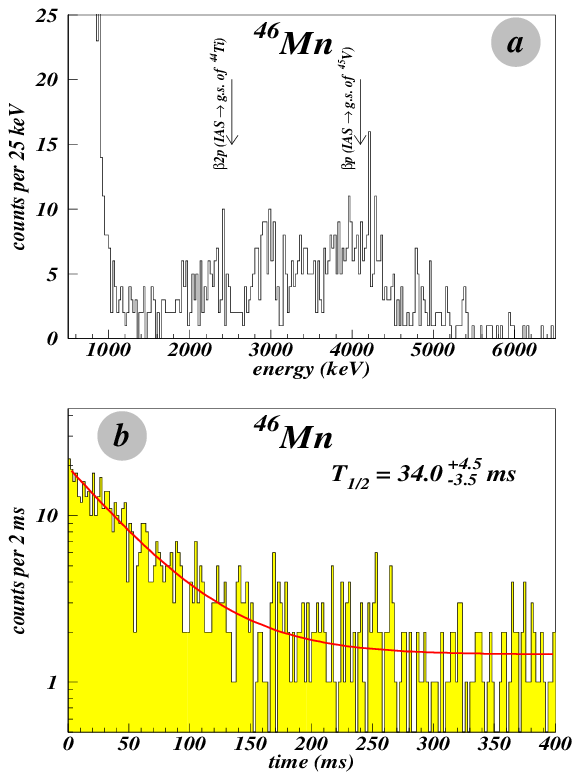}{$a)$ Decay-energy distribution of
           \noy{46}{Mn}. The arrows indicate the expected
           positions of charged-particle groups from $\beta p$ and
           $\beta 2p$ transitions via
           the IAS to the ground states of \noy{45}{V} and \noy{44}{Ti}.
           $b)$ Decay-time distribution of events after
           implantation of \noy{46}{Mn}. The half-life deduced is 
           $T_{1/2} = 34^{+4.5}_{-3.5}\ $ms.}

For this isotope, the half-life is determined to be
$T_{1/2} = 34.0^{+4.5}_{-3.5}\ $ms.
No subsequent decay by proton emission is expected
to contaminate the decay-energy distribution ab\-ove 1 MeV.
This result is in agreement with a previous result
from Borrel {\it et al.}~\cite{borrel92} (see table~\ref{periode}).
From the trigger efficiency and the number of counts in the
decay-energy distribution, we deduce a total proton branching ratio
of $P_p = (58\pm 9)\%$.

\subsection{The decay of ${}^{47}$Fe}

In figure~\ref{fe47_all}a, we present our results for the charged-particle
emission from \noy{47}{Fe}. We will analyse this spectrum together with
the coincident $\gamma$-ray spectrum shown in figure~~\ref{fe47_gamma}.
According to the structure of the mirror nucleus of 
the $\beta$-decay daughter \noy{47}{Mn},
we suppose that the transition at $E_p = (5975\pm 25)\ $keV
(see fig.~\ref{fe47_all}a) corresponds
to the delayed proton emission from the IAS in \noy{47}{Mn} 
to the first $2^+$ state in \noy{46}{Cr}.
This assumption is supported by the coincident observation of a $\gamma$
line at ($892\pm 1)\ $keV (see fig.~\ref{fe47_gamma}a) which most probably
correspond to the $2^+ \rightarrow 0^+$ transition in \noy{46}{Cr}.
In the mirror nucleus \noy{46}{Ti}, the $2^+$ state has an excitation energy of 
889.3~keV.
Using the ground-state mass excess of \noy{46}{Cr}, we deduce an excitation 
energy of the IAS of $E^* = (6867\pm 30 \pm 160)\ $keV, where the second 
error bar is due the $^{47}$Mn mass excess uncertainty.

\placefigure{8}{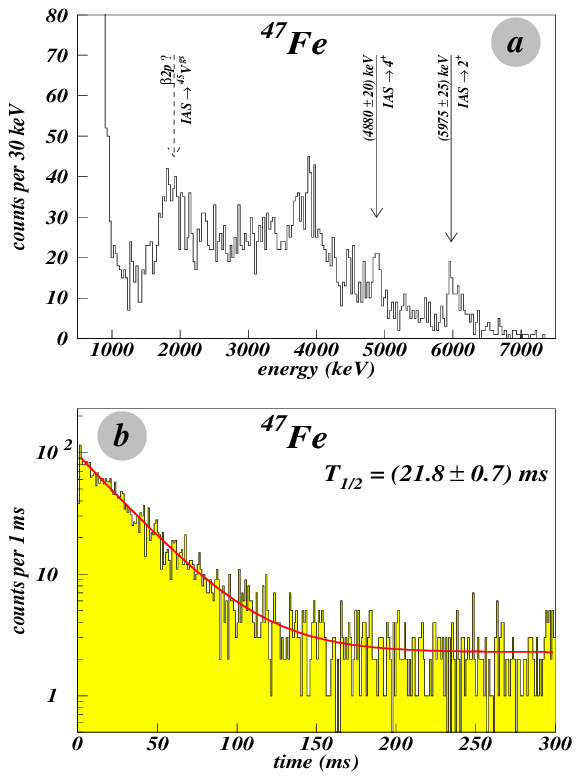}{$a)$ Decay-energy spectrum of \noy{47}{Fe}.
        The arrows indicate the charged-particle groups from the decay of 
        the IAS to the first and second excited state in \noy{46}{Cr}.
        $b)$ Decay-time distribution of decay events after
        implantation of \noy{47}{Fe}. The half-life deduced is 
        $T_{1/2} = (21.8\pm 0.7)\ $ms.}

In agreement with this excitation energy, we can assign the 
($4880\pm 20)\ $keV charged-particle peak to
the transition from the IAS to the first $4^+$ state in \noy{46}{Cr}.
In the $\gamma$-ray spectrum (see fig.~\ref{fe47_gamma}b), 
the $1095\ $keV line (1120.5~keV in the mirror nucleus), 
in coincidence with the $4880\ $keV peak, 
corresponds then to the $4^+ \rightarrow 2^+$ transition.
This assignment leads to an excitation energy of the IAS of
$E^* = (6869\pm 25 \pm 160)\ $keV which is in nice agreement with
the result above.
We deduce a mean value for the position of this IAS of
$E^* = (6868\pm 35 \pm 160)\ $keV.
The observation of the two $\gamma$ rays belonging to the ground-state 
rotational band of the even-even nucleus \noy{46}{Cr} is in agreement 
with what is expected from mirror symmetry.

\placefigure{8}{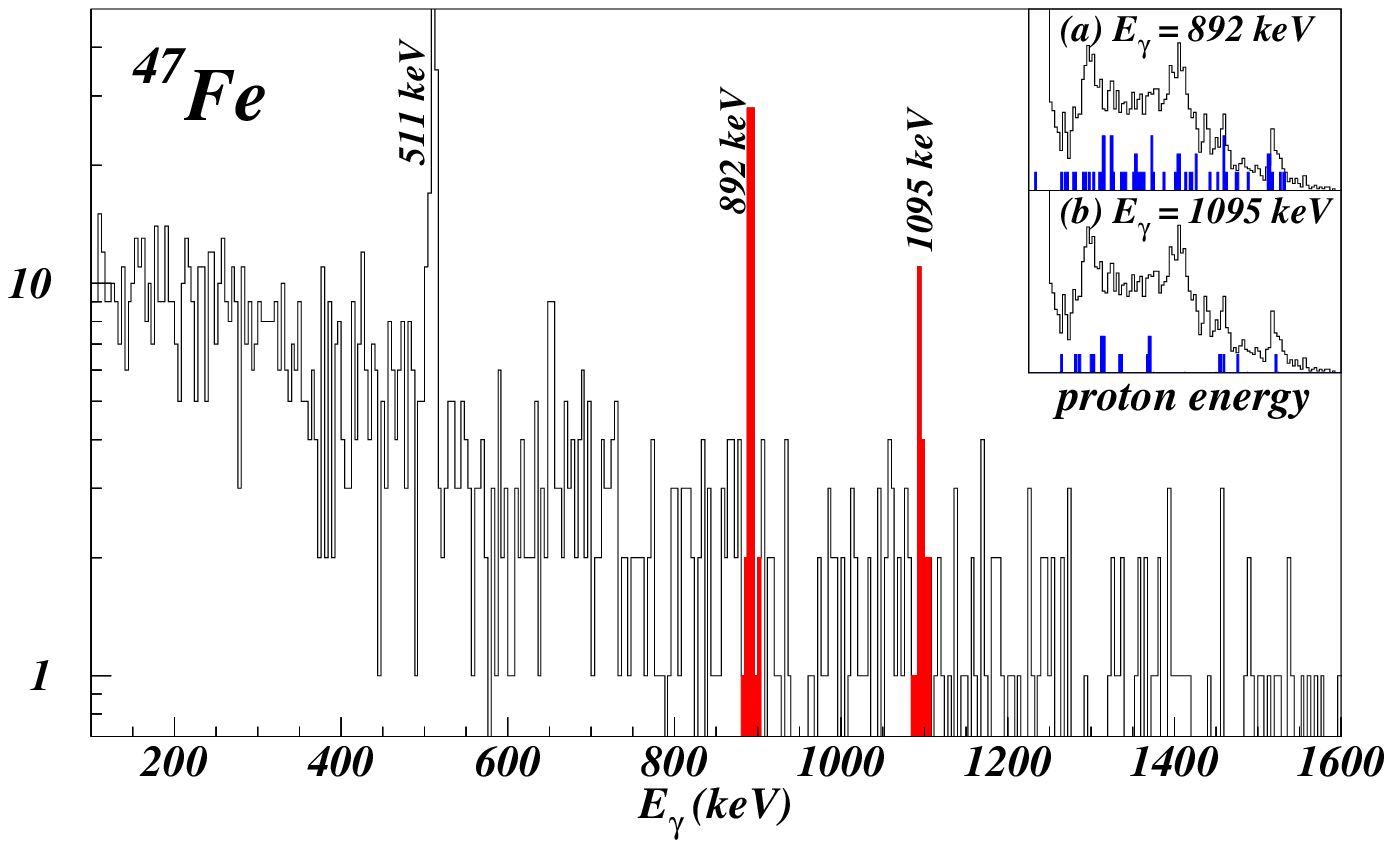}{$\gamma$-ray energies measured in coincidence 
        with protons. Inserts $a$ and $b$ show the distribution 
        of charged-particle events in coincidence with the $892$ 
        and $1095\ $keV $\gamma$ lines in comparison to
        the complete decay-energy distribution.}

According to the structure of the mirror nucleus of \noy{46}{Cr}, 
the strong proton transition around ($3890\pm 25)\ $keV may correspond
to the decay of the IAS to states around $3\ $MeV excitation energy, where
states with spin $J^\pi=2^+,3^-,4^-$ are expected which could be fed
by proton emission with low angular momentum ($\ell = 0,1$) from the IAS.
However, this strength can also result from proton emission
by low-lying states in \noy{47}{Mn} populated by Gamow-Teller decays
of \noy{47}{Fe}.

The $\beta 2p$ transition via the IAS ($J^\pi = 7/2^-$) to the
ground state of \noy{45}{V} is expected around $1915\ $keV.
Some proton strength is observed at this energy, but no single peak
can be clearly resolved. According to calculation by Detraz~\cite{detraz90}
the ratio between $\beta$2p and $\beta$p emission from the IAS is rather low 
(less than 0.05). This prediction is in agreement with the absence of a 
strong $\beta$2p peak in our experimental spectrum.

\placefigure{10}{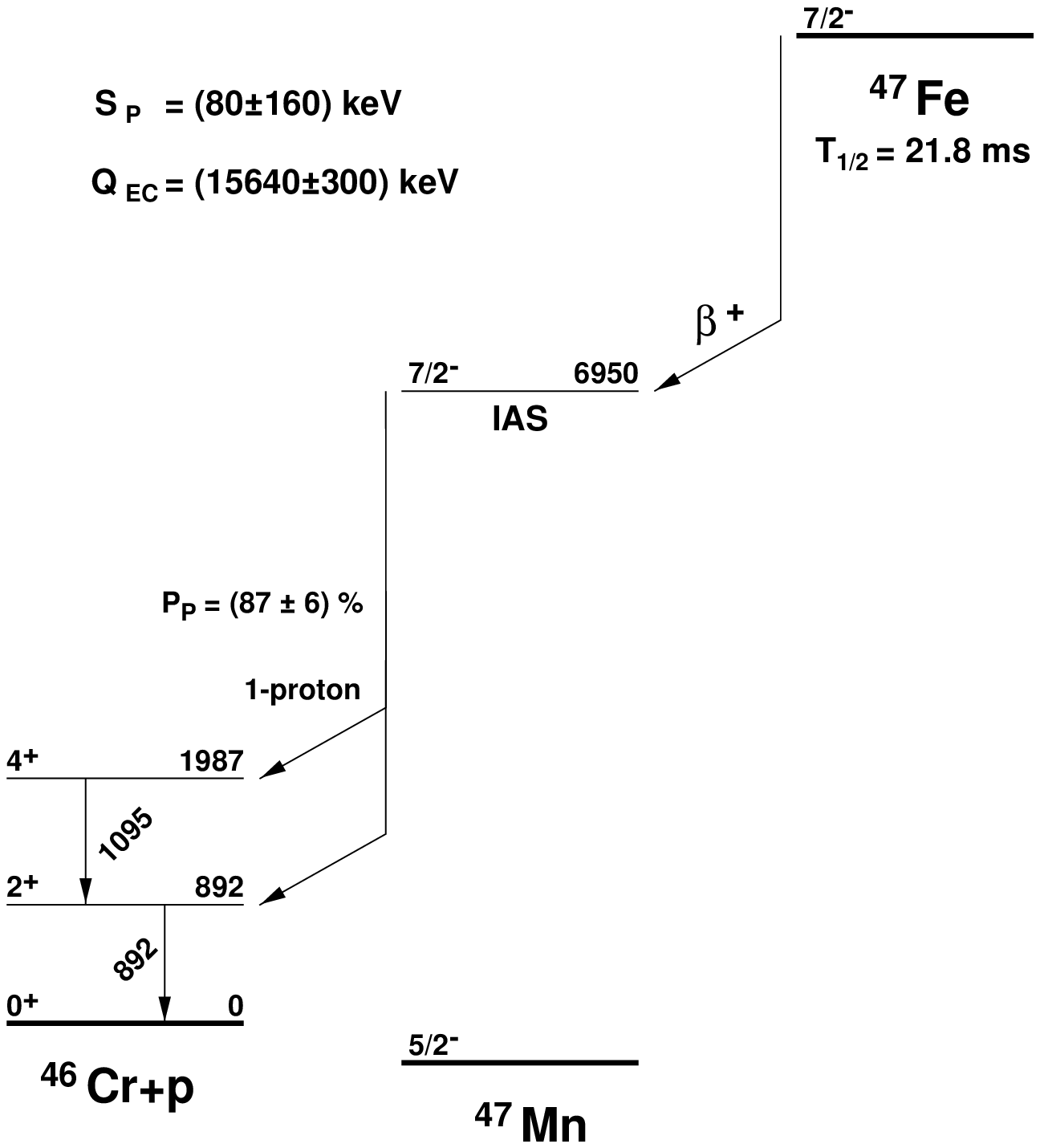}{Partial decay scheme for \noy{47}{Fe}. 
                 Energies on the scheme are given in keV.}

All protons in the charged-particle spectrum of figure~\ref{fe47_all}a 
are due to the decay of \noy{47}{Fe}. We deduce the half-life in the usual 
way and obtained a value of $T_{1/2} = (21.8\pm 0.7)\ $ms which is in 
agreement with but more precise than a previous measurement~\cite{borrel92}.
As a final result, we determine the proton branching ratio to be 
$P_p = (87\pm 7)\%$ for protons above 1~MeV.
The resulting decay scheme is shown in figure~\ref{fe47_decay}.

\subsection{The decay of ${}^{46}$Fe}

For the A=46, T=3 nuclei, three members of the isobaric mass multiplet
are probably known. We will use therefore the ground state of \noy{46}{Ca}, 
the 5.021~MeV state in \noy{46}{Sc}, and the 14.153~MeV state in \noy{46}{Ti}. 
For the IMME, we obtain 
the following mass excess values: $(-43.135 \pm 3)\ $MeV (\noy{46}{Ca}), 
$(-36.738 \pm 0.006)\ $MeV (\noy{46}{Sc}), and  $(-29.972 \pm 0.007)\ $MeV 
(\noy{46}{Ti}). With these data, we get the mass
excess values for \noy{46}{Fe} and for the IAS in \noy{46}{Mn} of
$(0.769 \pm 0.115)\ $MeV and $(-7.468 \pm 0.072)\ $MeV, respectively.
There is a slight uncertainty concerning the state in \noy{46}{Ti}, 
because there is a neighboring state at 14.300~MeV state with a 
tentative assignment of $I^\pi = 0^+$. However, whereas the 14.153~MeV
state yields parameters close to those expected from a uniformly
charged sphere~\cite{benenson79} and a \noy{46}{Fe} ground state mass
close to the mass evaluation of Audi and Wapstra~\cite{audi97}, the 14.300~MeV
state gives values far away from both criteria. 
Using these mass excess values, we expect the $\beta$p transition
via the IAS to the ground state of \noy{45}{Cr} at about 4.65~MeV in the 
decay-energy spectrum. The $\beta$2p transition to the ground state of
\noy{44}{V} is expected at 1.8~MeV.

Figure~\ref{fe46_all}a shows the charged-particle spectrum from the decay of 
\noy{46}{Fe}. No peak is seen for the $\beta$2p transition to the ground state. 
This is in agreement with the calculations of Detraz~\cite{detraz90} who predicts
a ratio between $\beta$2p and $\beta$p of less than 10\%. 

Although we clearly have seen several charged-particle lines in the
decay-energy spectrum of figure~\ref{fe46_all}a, we can not place any 
of them in a decay scheme. Proton-$\gamma$ coincidences are necessary to
identify the transitions involved.

The $\beta$p decay of \noy{46}{Fe} feeds states in \noy{45}{Cr}. The decay of 
this isotope has been studied by Jackson {\it et al.}~\cite{jackson74} who 
identified a delayed proton branch with an energy of 2.1~MeV. This peak
is visible in our spectrum. From the intensity of this peak and the 
known branching ratio for this 2.1~MeV proton group, we deduce
a $\beta$p branching ratio of $(36 \pm 20)\ \%$ for $^{46}$Fe. As the expected 
$\beta$2p branching ratio is much smaller~\cite{detraz90}, 
more than 50\% of the decay
of \noy{46}{Fe} feeds proton-bound levels in \noy{46}{Mn}. For a rather 
exotic isotope like \noy{46}{Fe} with a $\beta$-decay daughter having
a one-proton separation energy of only about 300~keV, a 50\%
feeding of $\gamma$-decaying levels or of the ground state of $^{46}$Mn seems to be rather high.
From the mirror nucleus, only one $I^\pi = 1^+$ level is expected in
\noy{46}{Mn} at around 1~MeV which might not decay by proton emission.
Using the mass excess of \noy{46}{Fe} as determined in the preceeding paragraph,
the mass excess of \noy{46}{Mn}~\cite{audi97}, and the half-life of \noy{46}{Fe}
(see below), a 50\% branching ratio to a state at about 1~MeV corresponds
to a $\log ft$ value of 3.2, much too small for a Gamow-Teller transition.
A possible explanation for this discrepancy might be a much larger 
$\beta$2p strength. However, higher-statistics data are clearly needed to
get a consistent picture for the decay of \noy{46}{Fe}.

Taking into account the daughter decay branches in \noy{45}{Cr} and \noy{46}{Mn}, 
the resulting half-life for \noy{46}{Fe} is $T_{1/2} = 12.0^{+4.2}_{-3.2}\ $ms 
(fig.~\ref{fe46_all}b)
where the uncertainty is the quadratic sum of uncertainties coming from
the fit, the daughter half-lives and the branching ratios.

\placefigure{8}{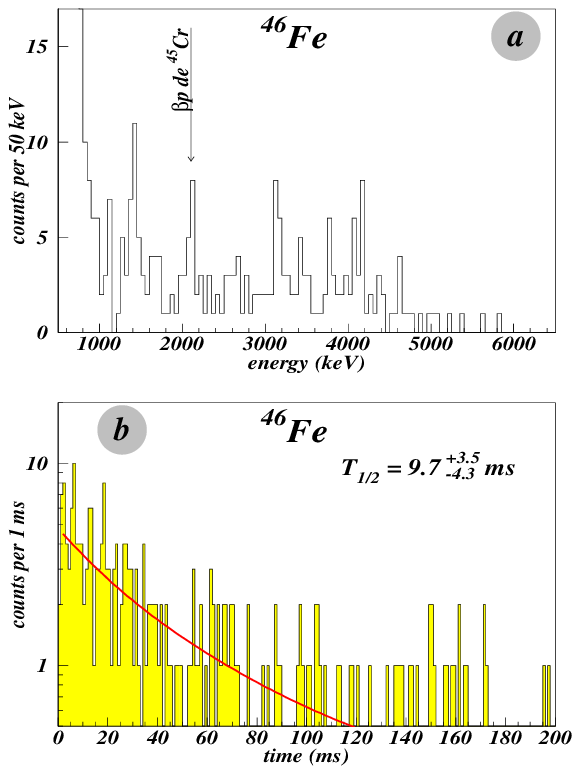}{$a)$ Decay-energy distribution of
    \noy{46}{Fe}. The peak from the daughter decay of \noy{45}{Cr} at
    2.1 MeV is indicated. 
    $b)$ Decay-time distribution of events after
    implantation of \noy{46}{Fe}. The resulting half-life is
    $T_{1/2} = 9.7^{+3.5}_{-4.3}\ $ms.}

\subsection{The decay of ${}^{45}$Fe}

According to theoretical predictions~\cite{brown91,ormand97,cole96}, 
\noy{45}{Fe} together with \noy{48}{Ni} is the best candidate for
two-proton radioactivity, i.e. a correlated emission of two protons 
from the ground state.
In such a case, due to their low energy, the protons may not exit the
implantation detector and we would measure only the sum energy of the decay.
However, since the acquisition was trigged only by events
in the neighbouring detector, these 2p events are lost and we should, in the 
case of a 2p radioactivity, only
observe the $\beta p$ decay of the daughter nucleus \noy{43}{Cr}.

In the decay-energy spectra of figure~\ref{fe45_all}a,b, we distinguish
again fast decays which occur within 12~ms after a \noy{45}{Fe}
implantation and events occuring in a time interval up to 150~ms which
should favour the observation of the daughter and grand-daughter decay.
However, the very poor statistics prevents any conclusion based on this 
spectrum. Charged-particle groups would be expected at about 
11.3, 8.5 and 8.3 MeV in the case of $\beta$p, $\beta$2p, or $\beta$3p
transition via the IAS to the respective ground states. No counts are 
observed in these regions due to the fact that i) the proton emission might
involve excited states in the grand-daughter nuclei and ii) the proton peak
efficiency decreases drastically for these high proton energies.

\placefigure{8}{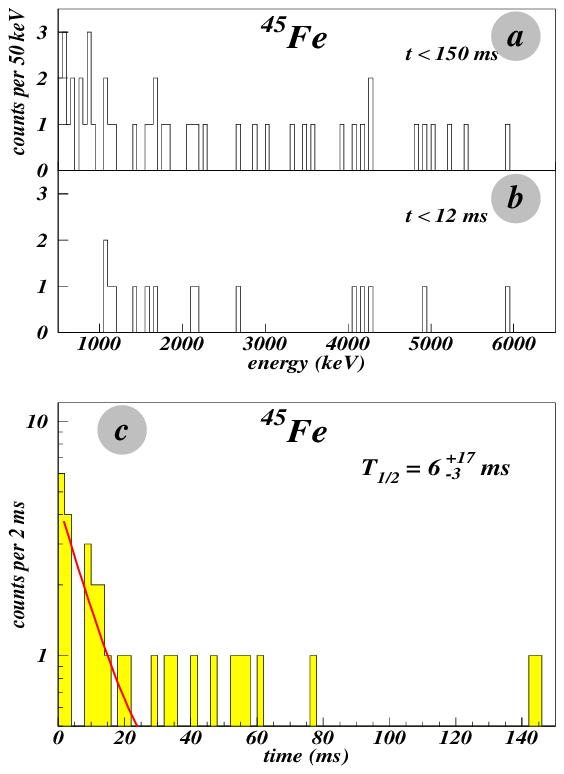}{$a,b)$ Decay-energy spectra  of \noy{45}{Fe}.
                 The histogram in part $a)$ is due to events up to
                 150~ms after a \noy{45}{Fe} implantation, whereas the 
                 spectrum in $b)$ contains decay events occuring less than 
                 12~ms after the implantation.
                 $c)$ Decay-time distribution of events after
                 implantation of \noy{45}{Fe} yielding a half-life of
                 $T_{1/2} = 6^{+17}_{-3}\ $ms.}

The most valuable information is probably contained in the decay-time
distribution shown in figure~\ref{fe45_all}c. The value is determined from
a fit taking into account different possibilities for the daughter decay.
The only contamination of the half-life determination comes from the
daughter decay of \noy{44}{Cr}. However, as the decay characteristics of this
nucleus are unknown, we assume that it might decay with a branching ratio between 
0 and 20\% by delayed-proton emission. This hypothesis is based on the
decay properties of \noy{52}{Ni}~\cite{faux96} and \noy{48}{Fe}~\cite{faux94}.
Our fitting procedure yields a half-life of $T_{1/2} = 6^{+17}_{-3}$ms (fig.~\ref{fe45_all}c).
This rather rough estimate of the half-life may be compared to theoretical
$\beta$-decay half-lives as shown in table~\ref{periode}. The theoretical
values are compatible with our experimental datum and indicate therefore
that $^{45}$Fe might decay mainly by $\beta$ decay and not by 
two-proton radioactivity. However,
higher-statistics data are clearly needed before concluding on the main
decay mode of this isotope.

\subsection{The decay of ${}^{49}$Ni}

Like \noy{42}{Cr}, \noy{49}{Ni} is a possible candidate for 2p radioactivity.
However, similar to \noy{42}{Cr}, theoretical calculations predict that
this isotope should decay mainly or even exclusively by $\beta$-delayed
charged-particle channels.

Figures~\ref{ni49_all}a and \ref{ni49_all}b present the decay-energy and -time
distribution for \noy{49}{Ni}. Although $\beta$-delayed charged-particle 
emission from the IAS in \noy{49}{Co} is energetically allowed 
($Q_{p} \sim 10.5$, $Q_{2p} \sim 7.3$, and $Q_{3p} \sim 7.2\ $MeV~\cite{audi97}), 
only one transition is clearly identified in the energy spectrum. The six
events at around 3.7~MeV are clearly not due to 2p radioactivity, because
in such a case the barrier-penetration half-life would be as short
as 10$^{-16}s$, many orders of magnitude smaller than the flight time
through the LISE3 separator. This peak is rather due to a $\beta$-delayed 
charged-particle branch. However, due to the poor statistics for 
this nucleus, no coincident $\gamma$ rays were observed. Therefore, we can 
not attribute the 3.7~MeV line to any well-defined transition.

\placefigure{8}{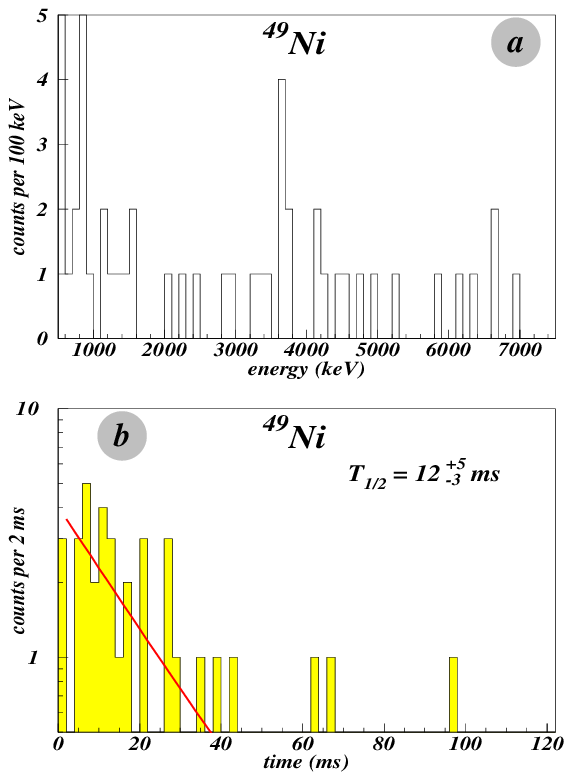}{$a)$ Decay-energy spectrum of
            \noy{49}{Ni}. The peak at 3.7~MeV is identified as being due to
            delayed charged-particle emission (see text).
            $b)$ Decay-time distribution of events after
            implantation of \noy{49}{Ni} yielding a half-life of
            $T_{1/2} = 12^{+5}_{-3}\ $ms.}

To estimate the half-life of \noy{49}{Ni}, our experimental results
do not allow to discriminate the different possible decay channels.
D\'etraz~\cite{detraz91} calculated a $\beta 2p/\beta p$ ratio of
3 for this nucleus. If the decay proceeds by delayed one-proton emission 
to \noy{48}{Fe}, we expected a subsequent $\beta p$ emission in less than 
20\% of the decays~\cite{faux96}.
The $\beta 2p$ or $\beta 3p$ channels are only followed by very weak 
delayed-particle emission channels.
Therefore, taking into account a possible contamination from the $\beta p$ 
decay of \noy{48}{Fe} in 0 to 10\% of the decays of \noy{49}{Ni},
we obtain a half-life of $T_{1/2} = 12^{+5}_{-3}\ $ms.
This value is in reasonable agreement with $\beta$-decay half-life
calculations (see table~\ref{periode}). From this comparison we conclude, 
as from the decay-energy spectrum, that \noy{49}{Ni} most probably decays
by $\beta$ emission.

\begin{table*}
\centerline{\vbox{\offinterlineskip
\halign{\tvi\tv\ccc{#}\tv&\ccc{#}&\ccc{#}&\ccc{#}\tv&\ccc{#}&\ccc{#}\tv\cr
\noalign{\hrule}
isotope &\multispan3{\ccc{theoretical prediction}}\tv&\multispan2{\ccc{experimental half-life}}\tv\cr
        &Gross Theory~\cite{tachibana90} &Ormand~\cite{ormand97} &
        Hirsch {\it et al.}~\cite{hirsch93} &previous work & this work \cr
\noalign{\hrule}
\noy{40}{Ti}&56     &56  &--        &$54\pm 2{}^{~(a)}$   &$53.5\pm 2.5$ \cr
 &      &   &  &  $52.7\pm 1.5{}^{~(b)}$ & \cr
\noy{39}{Ti}&28     &29  &--        &$26^{+8}_{-7}{}^{~(c)}$        &$31^{+6}_{-4}$ \cr
\noy{43}{Cr}&21     &14  &23 - 57   &$21\pm 4{}^{~(d)}$             &$21.6\pm 0.7$ \cr
\noy{42}{Cr}&21.2   &17  &12 - 33   &--                             & $13.4^{+3.6}_{-2.4}$ \cr
\noy{46}{Mn}&35     &37  &--        &$41^{+7}_{-6}{}^{~(d)}$        &$34.0^{+4.5}_{-3.5}$ \cr
\noy{47}{Fe}&27.3   &--  &22 - 57   &$27^{+32}_{-10}{}^{~(d)}$      &$21.8\pm 0.7$   \cr
\noy{46}{Fe}&24.3   &18  &15 - 30   &$20^{+20}_{-8}{}^{~(d)}$       &$9.7^{+3.5}_{-4.3}$ \cr
\noy{45}{Fe}&11.7   &\07 &\07 - 30  &--                             &$6.0^{+17}_{-3}$     \cr
\noy{49}{Ni}&10.7   &--  &\07 - 20  &--                             &$12^{+5}_{-3}$  \cr
\noalign{\hrule}
\noalign{\small ${}^{(a)}$ from W. Liu et al.~\cite{liu98}}
\noalign{\small ${}^{(b)}$ from W. Bhattacharya et al.~\cite{bhattacharya98}}
\noalign{\small ${}^{(c)}$ from C. Detraz et al.~\cite{detraz90}}
\noalign{\small ${}^{(d)}$ from V. Borrel et al.~\cite{borrel92}}
\noalign{\medskip}
}}}
\caption{Comparison of theoretical predictions for $\beta$-decay half-lives
         to results from previous experiments and from the present work. All
         values are given in ms.}
\label{periode}
\end{table*}

%
%
%
\section{Concluding remarks}

The present experiment performed at the GANIL/LISE3 facility 
using the fragmentation of a \noy{58}{Ni} primary beam yielded
valuable spectroscopic information concerning the decay of nuclei
in the region from titanium to nickel very close to or even beyond 
the proton drip-line. First half-life values have been determined for \noy{42}{Cr},
\noy{45}{Fe} and \noy{49}{Ni}.
In addition, higher-precision values have been obtained for \noy{39}{Ti},
\noy{43}{Cr}, \noy{46}{Mn} and \noy{46,47}{Fe}.

Although only few transitions could be clearly identified, delayed-proton
or two-proton emission was used to determine the excitation energy of the IAS
in the decay of \noy{43}{Cr} and \noy{47}{Fe}.
However, the precision obtained is strongly limited by the 
lack of precise ground-state mass excess values
of isotopes involved in the decays.
Despite the low detection efficiency, proton-$\gamma$ coincidences proved to be
a helpful tool to identify transitions in the decays of 
\noy{43}{Cr} and \noy{47}{Fe} and they give valuable
information on the decay scheme and the daughter-nucleus structure.

One of the purposes of decay studies in this mass region is to search for
two-proton radioactivity, since \noy{39}{Ti}, \noy{42}{Cr},
\noy{45}{Fe} and \noy{48,49}{Ni} are candidates to this decay process.
We did not observe this process. However, an improved detection setup
optimised for such a search, together with increased statistics should
allow to search for such a decay mode especially in the decay of
$^{45}$Fe, $^{48}$Ni, and $^{54}$Zn.

\section*{Acknowledgements}

We would like to acknowledge the continous effort of the GANIL accelerator 
staff to provide us with a stable, high-intensity beam. We 
express our sincere gratitude to the LISE staff for ensuring a smooth 
running of the LISE3 separator. This work was 
supported in part by the Polish Committee of Scientific Research under
grant KBN 2 P03B 036 15, the 
contract between IN2P3 and Poland, as well as by the Conseil R\'egional 
d'Aquitaine.

%
%


\begin{thebibliography}{10}

\bibitem{blank95kr}
B.~Blank, S.~Andriamonje, S.~Czajkowski, F.~Davi, R.~{Del Moral}, J.~P. Dufour,
  A.~Fleury, A.~Musqu{\`ere}, M.~S. Pravikoff, R.~Grzywacz, Z.~Janas,
  M.~Pf{\"u}tzner, A.~Grewe, A.~Heinz, A.~Junghans, M.~Lewitowicz, J.-E.
  Sauvestre, and C.~Donzaud {\em Phys. Rev. Lett. 74}, 4611 (1995).

\bibitem{blank00ni48}
B.~Blank, M.~Chartier, S.~Czajkowski, J.~Giovinazzo, M.~S. Pravikoff, J.-C.
  Thomas, G.~de~France, F.~de~Oliveira~Santos, M.~Lewitowicz, C.~Borcea,
  R.~Grzywacz, Z.~Janas, and M.~Pf{\"u}tzner {\em Phys. Rev. Lett. 84},
  1116, (2000)

\bibitem{cable83}
M.~D. Cable, J.~Honkanen, R.~F. Parry, S.~H. Zhou, Z.~Y. Zhou, and J.~Cerny
  {\em Phys. Rev. Lett. 50}, 404 (1983).

\bibitem{cable84}
M.~D. Cable, J.~Honkanen, E.~Schloemer, M.~Ahmed, J.~E. Reiff, Z.~Y. Zhou, and
  J.~Cerny {\em Phys. Rev. C 30}, 1276 (1984).

\bibitem{jahn85}
R.~Jahn, R.~L. McGrath, D.~M. Moltz, J.~Reiff, X.~J. Xu, J.~{\"A}yst{\"o}, and
  J.~Cerny {\em Phys. Rev. C 31}, 1576 (1985).

\bibitem{axelsson98}
L.~Axelsson, J.~{\"A}yst{\"o}, M.~Borge, L.~Fraile, H.~Fynbo, A.~Honkanen,
  P.~Hornshoj, A.~Jokinen, B.~Jonson, P.~Lipas, I.~Martel, J.~Mukha,
  T.~Nilsson, G.~Nyman, B.~Petersen, K.~Rissager, M.~Smedberg, and O.~Tengblad
  {\em Nucl. Phys. A 628}, 345 (1998).

\bibitem{borrel87}
V.~Borrel, J.~Jacmart, F.~Pougheon, A.~Richard, R.~Anne, D.~Bazin,
  H.~Delagrange, C.~D{\'e}traz, D.~Guillemaud-Mueller, A.~Mueller, E.~Roeckl,
  M.~Saint-Laurent, J.~Dufour, F.~Hubert, and M.~Pravikoff {\em Nucl. Phys. 
  A473}, 331 (1987).

\bibitem{aysto85}
J.~{\"A}yst{\"o}, D.~M. Moltz, X.~Xu, J.~Reiff, and J.~Cerny {\em Phys. Rev.
  Lett. 55}, 1384 (1985).

\bibitem{kryger95}
R.~A. Kryger, A.~Azhari, M.~Hellstr{\"o}m, J.~H. Kelley, T.~Kubo, R.~Pfaff,
  E.~Ramakrishnan, B.~M. Sherrill, M.~Thoennessen, S.~Yokoyama, R.~J. Charity,
  J.~Dempsey, A.~Kirov, N.~Robertson, D.~G. Sarantites, L.~G. Sobotka, and
  J.~A. Winger {\em Phys. Rev. Lett. 74}, 860 (1995).

\bibitem{bochkarev92}
O.~V. Bochkarev, A.~A. Korsheninnikov, E.~A. Kuz{'}min, I.~G. Mukha, L.~V.
  Chulkov, and G.~B. Yan{'}kov {\em Sov. J. Nucl. Phys. 55}, 955  (1992).

\bibitem{brown91}
B.~A. Brown {\em Phys. Rev. C 43}, R1513 (1991).

\bibitem{ormand97}
W.~E. Ormand {\em Phys. Rev. C 55}, 2407 (1997).

\bibitem{cole96}
B.~J. Cole {\em Phys. Rev. C 54}, 1240 (1996).

\bibitem{caurier94}
E.~Caurier, A.~P. Zuker, A.~Poves, and G.~Martinez-Pinedo {\em Phys. Rev. C 50}, 225 (1994).

\bibitem{bhattacharya98}
A.~G. M.~Bhattacharya, N.~I. Kaloskamis, E.~G. Adelberger, H.~E. Swanson,
  R.~Anne, M.~Lewitowicz, M.~G. Saint-Laurent, W.~Trinder, C.~Donzaud,
  D.~Guillemaud-Mueller, S.~Leenhardt, A.~C. Mueller, F.~Pougheon, and
  O.~Sorlin {\em Phys. Rev. C 58}, 3677 (1998).

\bibitem{liu98}
W.~Liu, M.~Hellstrom, R.~Collatz, J.~Benlliure, L.~Chulkov, D.~{Cortina Gil},
  F.~Farget, H.~Grawe, Z.~Hu, N.~Iwasa, M.~Pf{\"u}tzner, A.~Piechaczek,
  R.~Raabe, I.~Reusen, E.~Roeckl, G.~Vancraeynest, and A.~W{\"o}hr {\em Phys.
  Rev. C 58}, 2677 (1998).

\bibitem{detraz90}
C.~D{\'e}traz, R.~Anne, P.~Bricault, D.~Guillemaud-Mueller, M.~Lewitowicz,
  A.~C. Mueller, Y.~H. Zhang, V.~Borrel, J.~C. Jacmart, F.~Pougheon,
  A.~Richard, D.~Bazin, J.~P. Dufour, A.~Fleury, F.~Hubert, and M.~S. Pravikoff
  {\em Nucl. Phys. A519}, 529 (1990).

\bibitem{moltz92}
D.~M. Moltz, J.~C. Batchelder, T.~F. Lang, T.~J. Ognibene, J.~Cerny, P.~E.
  Haustein, and P.~L. Reeder {\em Z. Phys. A 342}, 273 (1992).

\bibitem{borrel92}
V.~Borrel, R.~Anne, D.~Bazin, C.~Borcea, G.~G. Chubarian, R.~{Del Moral},
  C.~D{\'e}traz, S.~Dogny, J.~P. Dufour, L.~Faux, A.~Fleury, L.~K. Fifield,
  D.~Guillemaud-Mueller, F.~Hubert, E.~Kashy, M.~Lewitowicz, C.~Marchand, A.~C.
  Mueller, F.~Pougheon, M.~S. Pravikoff, M.~G. Saint-Laurent, and O.~Sorlin
  {\em Z. Phys. A 344}, 135 (1992).

\bibitem{sissi}
A.~{Joubert {\em et al.}} {\em Proc. of the Second Conf. of the IEEE Particle
  Accelerator, San Francisco, May 1991}, 594.

\bibitem{lise}
A.~C. Mueller and R.~Anne {\em Nucl. Instrum. Meth. B56}, 559 (1991).

\bibitem{lise_code_new}
O.~{Tarasov et al.}, Proc. 5th Conf. Radioactve Nuclear Beams, Divonne,
  France, to be published \\ http://dnr080.jinr.ru/lise.html.

\bibitem{paw}
 CN/ASD Group, PAW Users Guide, CERN Program Library Q121, CERN (1993).

\bibitem{benenson79}
W.~Benenson and E.~Kashy {\em Rev. Mod. Phys. 51}, 527 (1979).

\bibitem{ormand96}
W.~E. Ormand {\em Phys. Rev. C 53}, 214 (1996).

\bibitem{audi97}
G.~Audi and A.~H. Wapstra {\em Nucl. Phys. A625}, 1 (1997).

\bibitem{ratkowski75}
A.~Ratkowski {\em Nucl. Instrum. and Meth. 130}, 533 (1975).

\bibitem{masses88}
P.~Haustein {\em At. Data Nucl. Data Tab. 39}, 185 (1988).

\bibitem{moeller95}
P.~M{\"o}ller, J.~R. Nix, W.~D. Myers, and W.~J. Swiatecki {\em At. Data Nucl.
  Data Tab. 59}, 185 (1995).

\bibitem{duflo95}
J.~Duflo and A.~Zuker {\em Phys. Rev. C 52}, R23 (1995).

\bibitem{tachibana90}
T.~Tachibana, M.~Yamada, and Y.~Yoshida {\em Prog. Theor. Phys. 84},
  641 (1990).

\bibitem{giovinazzo00}
J.~Giovinazzo, P.~Dessagne, and C.~Mieh{\'e} {\em Nucl. Phys. A 674},
  394, (2000).

\bibitem{jackson74}
K.~P. Jackson, J.~C. Hardy, H.~Schmeing, R.~L. Graham, J.~S. Geiger, and K.~W.
  Allen {\em Phys. Lett. B 49}, 341 (1974).

\bibitem{faux96}
L.~Faux, S.~Andriamonje, B.~Blank, R.~{Del Moral}, J.~P. Dufour, A.~Fleury,
  T.~Josso, M.~S. Pravikoff, S.~Czajkowski, Z.~Janas, A.~Piechaczek, E.~Roeckl,
  K.-H. Schmidt, K.~S{\"u}mmerer, W.~Trinder, M.~Weber, T.~Brohm, H.-G. Clerc,
  A.~Grewe, E.~Hanelt, A.~Heinz, A.~Junghans, C.~R{\"o}hl, S.~Steinh{\"a}user,
  B.~Voss, and M.~Pf{\"u}tzner {\em Nucl. Phys. A602}, 167 (1996).

\bibitem{faux94}
L.~Faux, M.~S. Pravikoff, S.~Andriamonje, B.~Blank, R.~{Del Moral}, J.-P.
  Dufour, A.~Fleury, C.~Marchand, K.-H. Schmidt, K.~S{\"u}mmerer, T.~Brohm,
  H.-G. Clerc, A.~Grewe, E.~Hanelt, B.~Voss, and C.~Ziegler {\em Phys. Rev. 
  C 49}, 2440 (1994).

\bibitem{detraz91}
C.~D{\'e}traz {\em Z. Phys. A 340}, 227 (1991).

\bibitem{hirsch93}
M.~Hirsch, A.~Staudt, K.~Muto, and H.~V. Klapdor-Kleingrothaus {\em At. Data
  Nucl. Data Tables 53}, 165 (1993).

\end{thebibliography}
\end{document}